\documentclass[aps,prd,preprint,nofootinbib]{revtex4}
\usepackage{graphicx}
\usepackage{mathrsfs}
\usepackage{amsfonts}
\usepackage{epsfig}

\begin{document}

\title{$B_c$ Production in Higgs Boson Decays}

\author{Jun Jiang$^{1,2}$\footnote{jiangjun13b@mails.ucas.ac.cn}
and Cong-Feng Qiao$^{1,2,3}$\footnote{qiaocf@ucas.ac.cn, corresponding author}}

\affiliation{$^1$School of Physics, University of Chinese Academy of Sciences, YuQuan Road 19A, Beijing 100049, China}
\affiliation{$^2$Key Laboratory of Vacuum Physics, University of Chinese Academy of Sciences}
\affiliation{$^3$CAS Center for Excellence in Particle Physics, Beijing 100049, China}

\begin{abstract}

$B_c$ production rate in Higgs boson decays is evaluated in NRQCD framework. Given Higgs total decay width is about $4.20~MeV$ and the vector $B^*_c$ meson decays completely to the ground state, we find that the branching fraction of $B_c$ meson production in Higgs decays is $8.50\times10^{-4}$, where both leading QCD and QED contributions are included. This process is hence detectable
in the high-luminosity/energy LHC.  It is found that the coupling of $Hb\bar{b}$ dominate the processes, contributions from the triangle top quark loop and other couplings ($Hc\bar{c}$, $HWW$ and $HZZ$) are small. In confronting to the quarkonia production, we find that the fraction rate of $B_c$ production is more than an order of magnitude bigger than those of charmonium and bottomonium production in Higgs decays. Moreover, various uncertainties and differential distributions of the concerned processes are analysed carefully.

\vspace {7mm} \noindent {\bf PACS number(s):}12.38.Bx, 14.40.Pq, 14.80.Bn

\end{abstract}

\maketitle

\section{Introduction}

In July 2012, the Higgs boson of the Standard Model has been found
by ATLAS \cite{Aad:2012tfa} and CMS \cite{Chatrchyan:2012xdj}
at Large Hardron Collider (LHC), which is a milestone in particle physics.
Recent results and review papers on Higgs measurements and searches at ATLAS and CMS can be found in Ref.s \cite{Orestano:2015ljx, Mochizuki:2015wua, Flechl:2015xxa, ac2015, Kourkoumelis:2015vza}.
Due to the insufficient Higgs samples and the detectors' limits, experimental study on the couplings of Higgs and fermions/bosons leaves much to be desired.

To open the era of precise Higgs physics, upgraded LHC, even new colliders are needed. High luminosity/energy scenarios are designed for LHC (HL/HE-LHC) \cite{Todesco:2013cca}.
Running at center-of-mass energy $\sqrt{s}=14~TeV$, cross-section of Higgs boson production at LHC
is about $55~pb$ (gluon-gluon fusion process dominates) \cite{higgscs}.
Given that the integrated luminosity is $3~ab^{-1}$, HL-LHC would produce $1.65\times10^8$ Higgs events.
While at HE-LHC who runs at $\sqrt{s}=33~TeV$, the cross-section of Higgs production would be about $200~pb$, hence Higgs events can be $3.5$ times bigger.
For the proposed Higgs factory, Circular Electron-Positron Collider (CEPC) running at $\sqrt{s}=250~GeV$ with an integrated luminosity of $5~ab^{-1}$, the cross-section of $e^+e^- \to H^0Z^0$ is about $0.219~pb$,
resulting in $1.10\times10^6$ Higgs events only \cite{CEPC-SPPCStudyGroup:2015csa}.
However at CEPC, since Higgs candidates can be identified through the recoil mass method without tagging its decays, Higgs production and decay are separated apart directly. Moreover at a $e^+e^-$ collider, physicists can perform measurements of model-independent Higgs total width and exclusive Higgs decay channels much better. The absolute values of Higgs coupling to bosons,
gluons and heavy fermions can also be measured. When updated to the Super Proton-Proton Collider (SPPC), researchers
can even measure the Higgs self-coupling, which is regarded as the holy grail of experimental particle physics.

With the excellent platforms, rare Higgs boson decay processes, like the heavy quarkonia production in Higgs decays, might be observed
for the first time. Pioneer investigation on the search of $H^0 \to J/\psi\gamma$ and $H^0 \to \Upsilon(nS) \gamma$ has been carried out by ATLAS \cite{Aad:2015sda}.
Theoretically, some related calculations have been done \cite{Achasov:1991ms, Bodwin:2013gca, Koenig:2015pha, Qiao:1998kv}.
Within the Non-Relativistic QCD (NRQCD) formulism \cite{Bodwin:1994jh,Petrelli:1997ge} and
light-cone methods \cite{Lepage:1980fj, Chernyak:1983ej}, both direct and indirect production mechanism and relativistic corrections to $H^0 \to J/\psi\gamma$ and $H^0 \to \Upsilon(nS) \gamma$ are studied \cite{Bodwin:2013gca}.
In addition, a detailed and complete analysis of $H^0 \to (\rho,\phi,\omega,J/\psi,\Upsilon(nS))+\gamma$ at the one-loop level is performed carefully \cite{Koenig:2015pha}.
Further, the complete inclusive production of heavy quarkonia ($J/\psi$ and $\Upsilon$) in Higgs boson decays are investigated in both color-singlet and color-octet mechanism \cite{Qiao:1998kv}.
Moreover, processes of Higgs boson decays to a quarkonium associated with a $Z^0$ boson are also analyzed \cite{Gao:2014xlv},
as well as double production of the quarkonia in Higgs decays \cite{Kartvelishvili:2008tz}.
Of course, all these rare decay channels might be observed and studied at the future HL/HE-LHC or CEPC/SPPC.

Containing two different flavors of heavy quarks, the $B_c$ mesons are very good research objets to reveal the nature of strong and weak interactions. Besides the study of direct one, indirect $B_c$ production through heavy particles decay are attractive.
It can inform us not only the nature $B_c$ itself, but also the properties of the parent particles.
Study of $B_c$ production through $W$ boson decays \cite{Qiao:2011yk, Liao:2011kd},
top quark decays \cite{Qiao:1996rd, Chang:2007si, Sun:2010rw} and $Z^0$ boson decays \cite{Chang:1992bb, Deng:2010aq, Qiao:2011zc, Jiang:2015jma} have been studied systematically and carefully.
Particularly, Ref. \cite{Sun:2010rw} and Ref.s \cite{Qiao:2011zc, Jiang:2015jma}
show us the precise next-to-leading order calculation of $B^{(*)}_c$ production,
which leads the calculation of $B^{(*)}_c$ production to the precise study level.

In this work, we will study the $B_c$ production in Higgs boson decays within NRQCD formulism, an estimation of events at HL/HE-LHC and a comparison with Higgs decays to the heavy quarkonia are also included.

The rest of this paper is organized as follows: Section II
presents our formalism and calculation method, numerical
evaluation and some discussion of the results are shown in Section III, and
Section IV gives a summary and some conclusions.

\section{Calculation Scheme Description and Formalism}
\begin{figure}[ht]
\centering
\includegraphics[width=10cm]{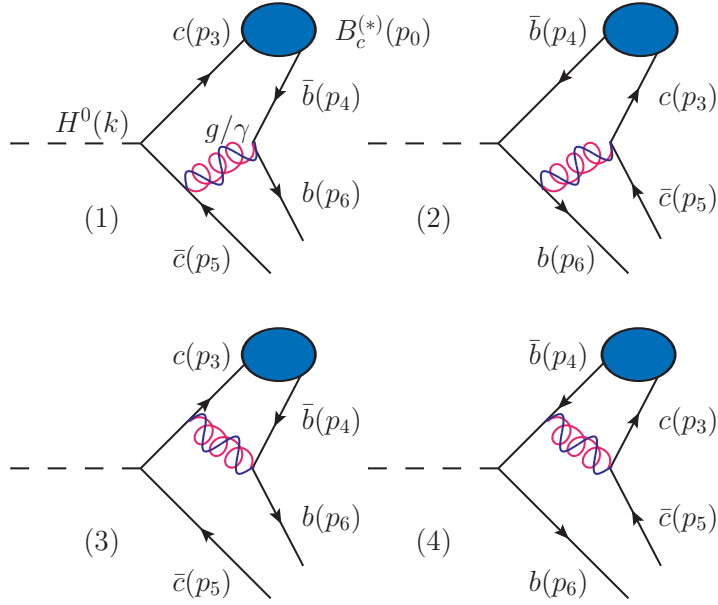}
\caption{\label{fig1} The QCD/QED Feynman diagrams of processes $H^0(k) \to B_c^{(*)}(p_0)+\bar{c}(p_5)+b(p_6)$,
where convolved/waved lines represent the gluon/photon propagators.}
\end{figure}

Feynman diagrams for processes $H^0(k) \to B_c^{(*)}(p_0)+\bar{c}(p_5)+b(p_6)$ are displayed in Fig.\ref{fig1}.
In our calculation, both QCD and QED contributions have been considered.
According to the NRQCD framework, the decay width can be factorized as
\begin{equation}
\Gamma=\sum_{n} \hat\Gamma_n(H^0 \to (c\bar{b})[n]+\bar{c}+b) \times \langle {\cal O}[n]\rangle \label{eq1}.
\end{equation}
In which the width $\hat\Gamma_n$ represents the short-distance coefficients at the partonic level which can be calculated perturbatively, long-distance matrix element $\langle{\cal O}[n]\rangle$ contains all the non-perturbative hardronization information, and $n$ stands for the involved $c\bar{b}$ Fock states.
In our computation, two color-singlet states $(c\bar b)[^1S_0]$ and $(c\bar b)[^3S_1]$
(donated as $B_c$ and $B^*_c$ respectively) are taken into consideration.
And their matrix elements can be related directly to the regularized Shr$\ddot{o}$dinger wave functions at the origin:
\begin{eqnarray}
\langle B_c|{\cal O}[^1S_0]| B_c \rangle \approx \frac{1}{4\pi}|\bar{R}_{B_c}|^2, \nonumber\\
\langle B^*_c|{\cal O}[^3S_1]| B^*_c \rangle \approx \frac{1}{4\pi}|\bar{R}_{B^*_c}|^2, \label{eq2}
\end{eqnarray}
where radial wave functions $\bar{R}_{B^{(*)}_c}$ can be computed by potential models \cite{Eichten:1995ch}. Note that the spin-splitting effect of the wave functions is ignored in our calculation, i.e. $|\bar{R}_{B_c}|=|\bar{R}_{B^*_c}|$.

For the partonic coefficient $\hat\Gamma_n$, its differential form can be expressed as
\begin{equation}
d\hat\Gamma_n = \frac{1}{2m_H}\sum |M(n)|^2 d\Phi_3, \label{eq3}
\end{equation}
where $\sum$ sums over the spin, color and polorization, $\Phi_3$ is the 3-body phase space of the final states,
$m_H$ is the mass of Higgs boson. The amplitude $M(n)$ which contains both QCD and QED contributions has the form of
\begin{equation}
M(n) = \sum_{i=1}^4 (\mathcal{C_{QCD}}+\mathcal{C_{QED}})\bar{u}(p_6)\cdot M^{\prime}_i(n)\cdot v(p_5),\label{eq4}
\end{equation}
where $\mathcal{C_{QCD}}=\frac{e C_F g_s^2}{2 m_W sin\theta_W}$ and $\mathcal{C_{QED}}=-\frac{e^3}{9 m_W sin\theta_W}$,
with $m_W$ is the mass of W boson, $sin\theta_W$ is the sine of Weinberg angle $\theta_W$.
And $M^{\prime}_i(n)$ can be read from Fig.\ref{fig1} directly with the help of
Mathematica package FeynArts \cite{Hahn:2000kx}, their analytical expressions are
\begin{eqnarray}
M^{\prime}_1(n) &=& \frac{m_c \gamma_{\nu} \cdot \Pi(n) \cdot (-\not\!p_4-\not\!p_5-\not\!p_6+m_c) \cdot \gamma^{\nu}}
{(p_4+p_6)^2((p_4+p_5+p_6)^2-m_c^2)}, \nonumber \\
M^{\prime}_2(n) &=& \frac{m_b \gamma_{\nu} \cdot (\not\!p_3+\not\!p_5+\not\!p_6+m_b) \cdot \Pi(n) \cdot \gamma^{\nu}}
{(p_3+p_5)^2((p_3+p_5+p_6)^2-m_b^2)}, \nonumber \\
M^{\prime}_3(n) &=& \frac{m_c \gamma_{\nu} \cdot \Pi(n) \cdot \gamma^{\nu} \cdot (\not\!p_3+\not\!p_4+\not\!p_6+m_c)}
{(p_4+p_6)^2((p_3+p_4+p_6)^2-m_c^2)}, \nonumber \\
M^{\prime}_4(n) &=& \frac{m_b (-\not\!p_3-\not\!p_4-\not\!p_5+m_b) \cdot \gamma_{\nu} \cdot \Pi(n) \cdot \gamma^{\nu}}
{(p_3+p_5)^2((p_3+p_4+p_5)^2-m_b^2)}.\label{eq5}
\end{eqnarray}
In which, $\Pi(n)$ is the projector of Fock state $n$, which has the form of \cite{Bodwin:2002hg}
\begin{equation}
\Pi(n) = \frac{1}{2\sqrt{m_c+m_b}}\epsilon(n)(\not\!p_0+m_c+m_b)\frac{\delta_{ij}}{\sqrt{N_c}},\label{eq6}
\end{equation}
where $\epsilon(^1S_0)=\gamma_5$ , $\epsilon(^3S_1)=\not\! \epsilon$ with $\epsilon^{\alpha}$ is
the polarization vector of $^3S_1$ state, $i$ and $j$ are the color indexes of the constitute quarks.

Submiting Eq.s (\ref{eq4}-\ref{eq6}) into Eq. (\ref{eq3}),
squared amplitude $|M(n)|^2$ can be obtained with the help of the Mathematica package FeynCalc \cite{Mertig:1990an}.
To construct the differential phase space $d\Phi_3$, we adopt partly the fortran codes of the package FormCalc \cite{Hahn:1998yk}.
And with its help, the kinematic and dynamic variables can be extracted out easily
to carry out the numerical analysis of differential distributions.

\section{Numerical Results and Analysis}

In the numerical computation, the one-loop running coupling constant is employed. And the values of radial wave functions can be found in Ref. \cite{Eichten:1995ch}. In our calculation, wave functions evaluated by QCD (Buchm$\ddot{u}$ller-Tye) potential model is adopted.
For convenience, some input parameters are listed as follows \cite{Agashe:2014kda}:
\begin{eqnarray}
&&m_c=1.5GeV, m_b=4.9GeV, m_H=125.7GeV, \nonumber \\
&&\alpha=1/127, \alpha_s(2m_c)=0.326, \alpha_s(2m_b)=0.214, \nonumber \\
&&m_W=80.4GeV, sin\theta_W=\sqrt{0.231}, m_{B^{(*)}_c}=m_b+m_c, \nonumber \\
&&m_Z=91.2GeV, V_{cb}=0.0414, m_t=173.2GeV. \label{eq7}
\end{eqnarray}

In Table \ref{tab1}, the QCD and QED contributions to decay widths $\Gamma(H^0 \to B_c^{(*)}+\bar{c}+b)$ are presented. In comparison with the QCD one, QED contribution is
negligible. According to the Feynman rules, the coupling of quarks and photons is related to the charge of quarks
while that of quarks and gluons is not,
which results in the negative contributions from their cross-terms. In the rest of the paper,
the decay widths mentioned refer to the total ones if there is no special instructions.
\begin{table}
\caption{\label{tab1} QCD and QED contributions to the decay widths $\Gamma(H^0 \to B_c^{(*)}+\bar{c}+b)$ (units: KeV).}
\begin{center}
  \begin{tabular}{c|c|c}
  \toprule
   & $~B_c~$ & $~B^*_c~$ \\
   \hline
	QCD & 1.53 & 2.08 \\
	QED & 5.71$\times 10^{-5}$ & 7.75$\times 10^{-5}$ \\
	cross-terms & -0.0187 & -0.0254 \\
	\hline
	total decay width & 1.51 & 2.06 \\
  \botrule
  \end{tabular}
\end{center}
\end{table}
\begin{table}
\caption{\label{tab2} Uncertainties of the decay widths $\Gamma(H^0 \to B_c^{(*)}+\bar{c}+b)$ (units: KeV) caused by the matrix element $\langle B^{(*)}_c |{\cal O}[^1S_0(^3S_1)]| B^{(*)}_c \rangle$ (units: $GeV^3$). Percentages in brackets are corrections relative to the QCD (B-T) one.}
\begin{center}
  \begin{tabular}{c|c|c}
  \toprule
   $\langle B^{(*)}_c |{\cal O}[^1S_0(^3S_1)]| B^{(*)}_c \rangle$ & $B_c$ & $B^*_c$ \\
    \hline
      QCD (B-T): 0.1306  & 1.51 & 2.06 \\
      Power-law: 0.1361 & 1.57(+4\%) & 2.15(+4\%) \\
      Logarithmic: 0.1200 & 1.39(-8\%) & 1.89(-8\%) \\
      Cornell: 0.2534 & 2.93 & 4.00 \\
  \botrule
  \end{tabular}
\end{center}
\end{table}

For our complete leading-order NRQCD calculation, the uncertainty sources mainly include the non-perturbative matrix element $\langle {\cal O}[n]\rangle$, the mass parameters $m_b$ and $m_c$, and the running coupling constant $\alpha_s(\mu)$.
To estimate the uncertainties from matrix element $\langle B^{(*)}_c |{\cal O}[^1S_0(^3S_1)]| B^{(*)}_c \rangle$, we adopt the four potential models in Ref. \cite{Eichten:1995ch}, and the corresponding decay widths are displayed in Table \ref{tab2}. If adopting the Power-law potential and Logarithmic potential as the upper and lower limits respectively, we obtain the corrections to the widths are about $+4\%$ and $-8\%$ accordingly for both $B_c$ and $B_c^*$ cases.
In Table \ref{tab3}, decay widths with varying quark masses are presented, in which when $m_b(m_c)$ varies, the $m_c(m_b)$ is fixed at its
central value. The uncertainty caused by varying $m_b$ is about $8\%$ and $10\%$ for $B_c$ and $B_c^*$ respectively, while that caused by varying $m_c$ is more than two times bigger.
By taking QED Feynman diagrams into account,
the decay widths now are no longer proportional to $\alpha_s^2(\mu)$ alone, but have the form of
\begin{eqnarray}
\Gamma_{B_c}(\alpha_s)&=& 33.4\alpha_s^2-0.0873\alpha_s+0.000057,\nonumber\\
\Gamma_{B^*_c}(\alpha_s)&=& 45.3\alpha_s^2-0.119\alpha_s+0.000077. \label{eq8}
\end{eqnarray}
Considering that the coupling constant $\alpha_s(\mu)$ decreases by $10\%$ from $\alpha_s(2m_b)$, then decay width
$\Gamma_{B_c}$ would decrease from $1.51~KeV$ down to $1.22~KeV$, and $\Gamma_{B^*_c}$ goes from $2.06~KeV$ down to $1.66~KeV$. Both have about $20\%$ depression.
\begin{table}
\caption{\label{tab3} Uncertainties of the decay widths $\Gamma(H^0 \to B_c^{(*)}+\bar{c}+b)$ (units: KeV) caused by the mass parameters $m_b$ and $m_c$ (units: GeV). When $m_b(m_c)$ varies, the $m_c(m_b)$ is fixed at its central value.}
\begin{center}
  \begin{tabular}{c|c|c}
  \toprule
   $ m_{b,c} $& $B_c$ & $B^*_c$ \\
    \hline
      $m_b=4.9^{+0.2}_{-0.2}$ & $1.51^{+0.12}_{-0.12}$ & $2.06^{+0.20}_{-0.19}$ \\
      $m_c=1.5^{+0.1}_{-0.1}$ & $1.51^{-0.27}_{+0.35}$ & $2.06^{-0.40}_{+0.53}$ \\
  \botrule
  \end{tabular}
\end{center}
\end{table}
\begin{figure}[ht]
\centering
\includegraphics[width=10cm]{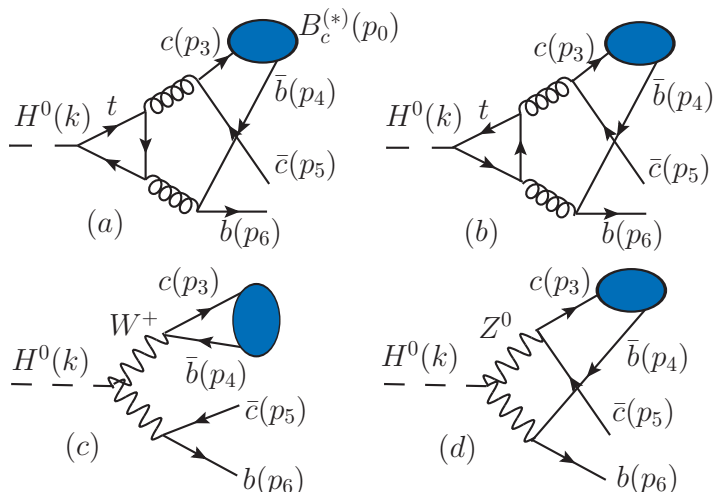}
\caption{\label{fig2} The triangle loop (top quark only) and W/Z boson propagated Feynman diagrams of processes $H^0(k) \to B_c^{(*)}(p_0)+\bar{c}(p_5)+b(p_6)$.}
\end{figure}
\begin{table}
\caption{\label{tab4} Corrections to the decay widths $\Gamma(H^0 \to B_c^{(*)}+\bar{c}+b)$ (units: KeV) raised from Fig. \ref{fig2}. Some input parameters are listed in Eq. (\ref{eq7}), $ \alpha_{ew} $ is the electroweak coupling constant and $ V_{cb} $ is the CKM matrix element for the mixing of $c,b$ quarks. Percentages in the brackets are the ratios of various corrections relative to the Fig. \ref{fig1} (QCD only) one. Note that only QCD Feynman diagrams in Fig. \ref{fig1} are considered.}
\begin{center}
  \begin{tabular}{c|c|c|c}
  \toprule
   Contributions & Order & $B_c$ & $B^*_c$ \\
    \hline
      Fig. \ref{fig1} (QCD only) & $\alpha_{ew} \alpha_s^2$ & 1.53 & 2.08  \\
      \hline
      cross-terms of Fig. (\ref{fig1} ,\ref{fig2}(a,b)) & $\alpha_{ew} \alpha_s^3$ & $5.32\times 10^{-2}$(+3.5\%) & $6.98\times 10^{-3}$(+0.34\%) \\
      Fig. \ref{fig2}(c) & $\alpha_{ew}^3 V_{cb}^4$ & $2.58\times 10^{-7}$(+0.0\%) & $2.67\times 10^{-7}$(+0.0\%) \\
      Fig. \ref{fig2}(d) & $\alpha_{ew}^3$ & $1.65\times 10^{-3}$(+0.11\%) & $1.67\times 10^{-3}$(+0.08\%) \\
      cross-terms of Fig. (\ref{fig1}, \ref{fig2}(c,d)) & --- & $-5.69\times 10^{-4}$(-0.037\%) & $8.23\times 10^{-6}$(+0.0\%) \\
  \botrule
  \end{tabular}
\end{center}
\end{table}

Now, let's discuss the uncalculated corrections of order-$v^2$, order-$ \alpha_s^3 $ and order-$ v^2\alpha_s^3 $, where $ v $ is the heavy quark or anti-quark velocity in the bound state rest frame. Generally, $ v^2\sim 25\% $ for $ J/\psi $ and $ v^2\sim 10\% $ for $ \Upsilon $.
Assuming $ v^2\sim 20\% $ for $ B^{(*)}_c $ and $ \alpha_s(2m_b)\sim 0.2 $, we estimate the order-$v^2$ correction to be $ 20\% $, order-$ \alpha_s^3 $ correction to be $ 20\% $ and order-$ v^2\alpha_s^3 $ correction to be $ 4\% $. In fact, we calculated the order-$ \alpha_s^3 $ contribution from the triangle loop (top quark only) Feynman diagrams in Fig. \ref{fig2}(a,b), whose correction to the decay widths $\Gamma(H^0 \to B_c^{(*)}+\bar{c}+b)$ is presented in Table \ref{tab4}. Where percentages in the brackets are the ratios of various corrections relative to the Fig. \ref{fig1} (QCD only) one. It is found that correction raised from the triangle top quark loop for $B_c$ case is bigger by a factor of 10 than that for $B^*_c$ one\footnote{This seems strange but we checked the calculation carefully.}. Corrections through the propagation of Higgs fragments to double W/Z bosons (Fig. \ref{fig2}(c,d)) are also included, all of which are quite small.

Assuming that the vector state $B^*_c$ decays to the ground state $B_c$
with 100\% efficiency through electromagnetic interaction, then at $\alpha_s(2m_b)=0.214$ we have the total decay width (Since the corrections listed in Table \ref{tab4} are small, data in Table \ref{tab1} are used only.):
\begin{eqnarray}
\Gamma_{total}({H^0 \to B_c+\bar{c}+b})=3.57(^{-0.67}_{+0.88})~KeV, \label{eq9}
\end{eqnarray}
where the errors are caused by varying $m_c$ only, which should be the biggest uncertainty source (as high as 25\% correction). If we adopt the Higgs total decay width
$\Gamma_{H}=4.20~MeV$ \cite{Heinemeyer:2013tqa}, then we obtain the total branching fraction:
\begin{eqnarray}
Br_{total}({H^0 \to B_c+\bar{c}+b})&=& \frac{\Gamma_{total}({H^0 \to B_c+\bar{c}+b})}{\Gamma_{H}}\nonumber \\
&=&8.50(^{-1.6}_{+2.1})\times10^{-4}. \label{eq10}
\end{eqnarray}

Presently experimental results on Higgs couplings have no deviation from the Standard Model predictions \cite{Orestano:2015ljx, Mochizuki:2015wua, Flechl:2015xxa, ac2015, Kourkoumelis:2015vza}, yet the new physics might lie in the Yukawa coupling highly possible. Presumably, the total branching fraction $Br_{total}$ has a strong dependence only on the $Hb\bar{b}$ couping\footnote{This is reasonable since the contributions from the triangle top quark loop and other couplings ($Hc\bar{c}$, $HZZ$ and $HWW$) are relatively small, see Tables (\ref{tab4}, \ref{tab5}).}. Given a adjustable factor $\kappa_b$ to the $Hb\bar{b}$ couping, we obtain a $\kappa_b^2$ factor to the total branching fraction. For example, 300\% correction ($\kappa_b^2=4$) to the total branching fraction implies a 100\% deviation from the Standard Model $Hb\bar{b}$ couping. In addition, new heavier particles which couple to Higgs boson could give an appreciable enhancement to the triangle loop in Fig. \ref{fig2}(a,b), and/or to the coupling similar to the ones in Fig. \ref{fig2}(c,d) if the new particles interact with gluons and quarks.

Running at $\sqrt{s}=14~TeV$ and with the integrated luminosity of $3~ab^{-1}$,
HL-LHC could produce $1.65\times10^8$ Higgs boson.
Hence we can obtain about $1.4\times10^5$ $B_c$ events according to Eq. (\ref{eq10}).
For the $B_c$ detection, the fully constructed channel is $B_c \to J/\psi(1S) \pi^+$,
whose branching fraction is about $0.5\%$ \cite{Chang:1992pt}.
Further considering that the branching rate of $J/\psi \to l^+l^-(l=e,\mu)$ is about 12\% \cite{Agashe:2014kda},
number of $B_c$ candidates produced in Higgs decays at HL-LHC is about $80$.
When center-of-mass energy is upgraded to $33~TeV$ (i.e. at HE-LHC), the $B_c$ events produced would be approaching $300$.
Then it might be a choice for experimental physicists to study the couplings of Higgs and heavy quarks through the Higgs decays to $B_c$ channel.
\begin{figure}
\centering
\includegraphics[width=8cm]{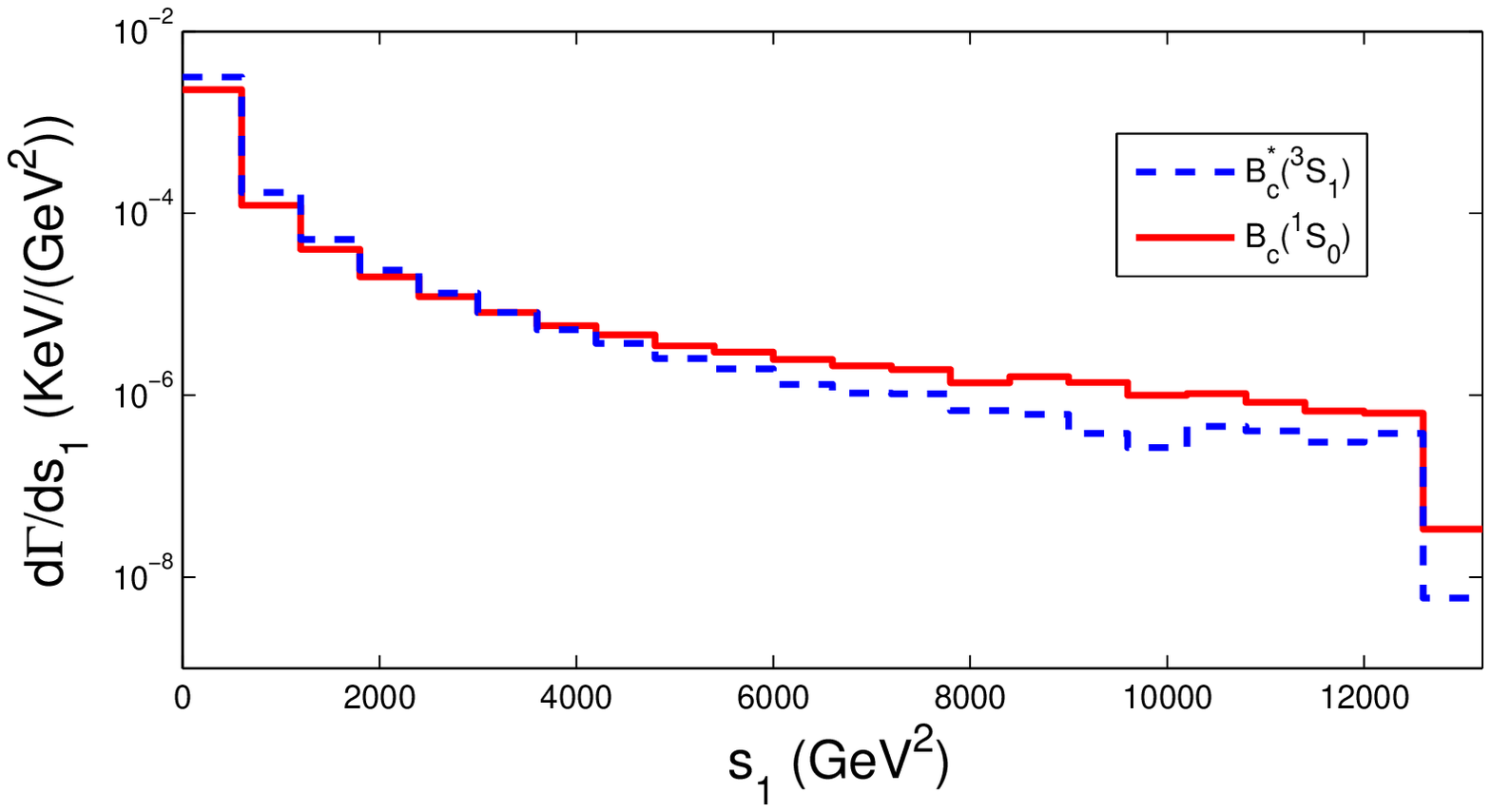}
\includegraphics[width=8cm]{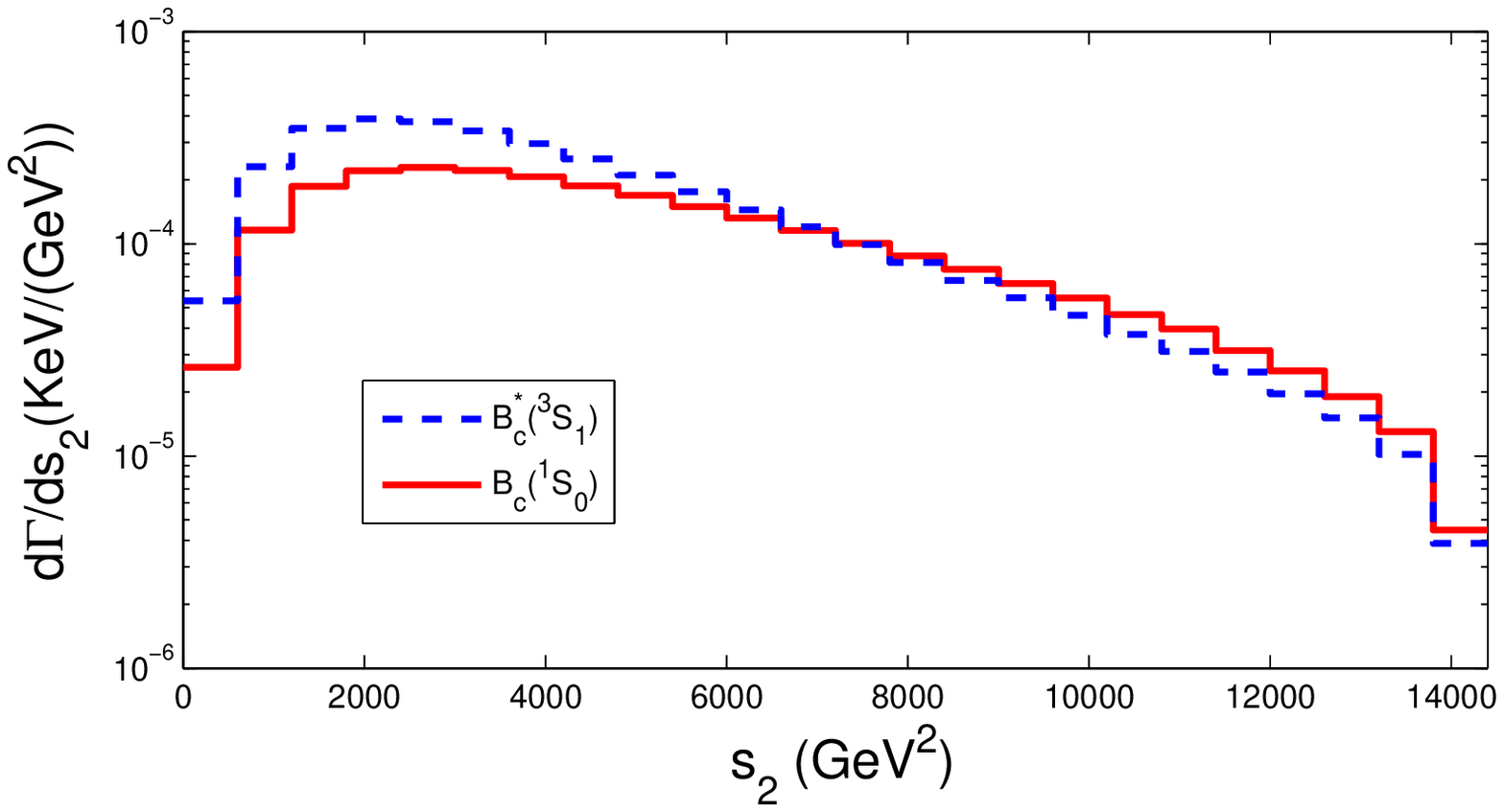}
\caption{\label{fig3} Differential decay widths $d\Gamma/ds_1$ (left) and $d\Gamma/ds_2$ (right) for processes $H^0(k) \to B_c^{(*)}(p_0)+\bar{c}(p_5)+b(p_6)$, where $s_1=(p_0+p_5)^2$ and $s_2=(p_5+p_6)^2$ are the invariant masses.}
\end{figure}
\begin{figure}
\centering
\includegraphics[width=8cm]{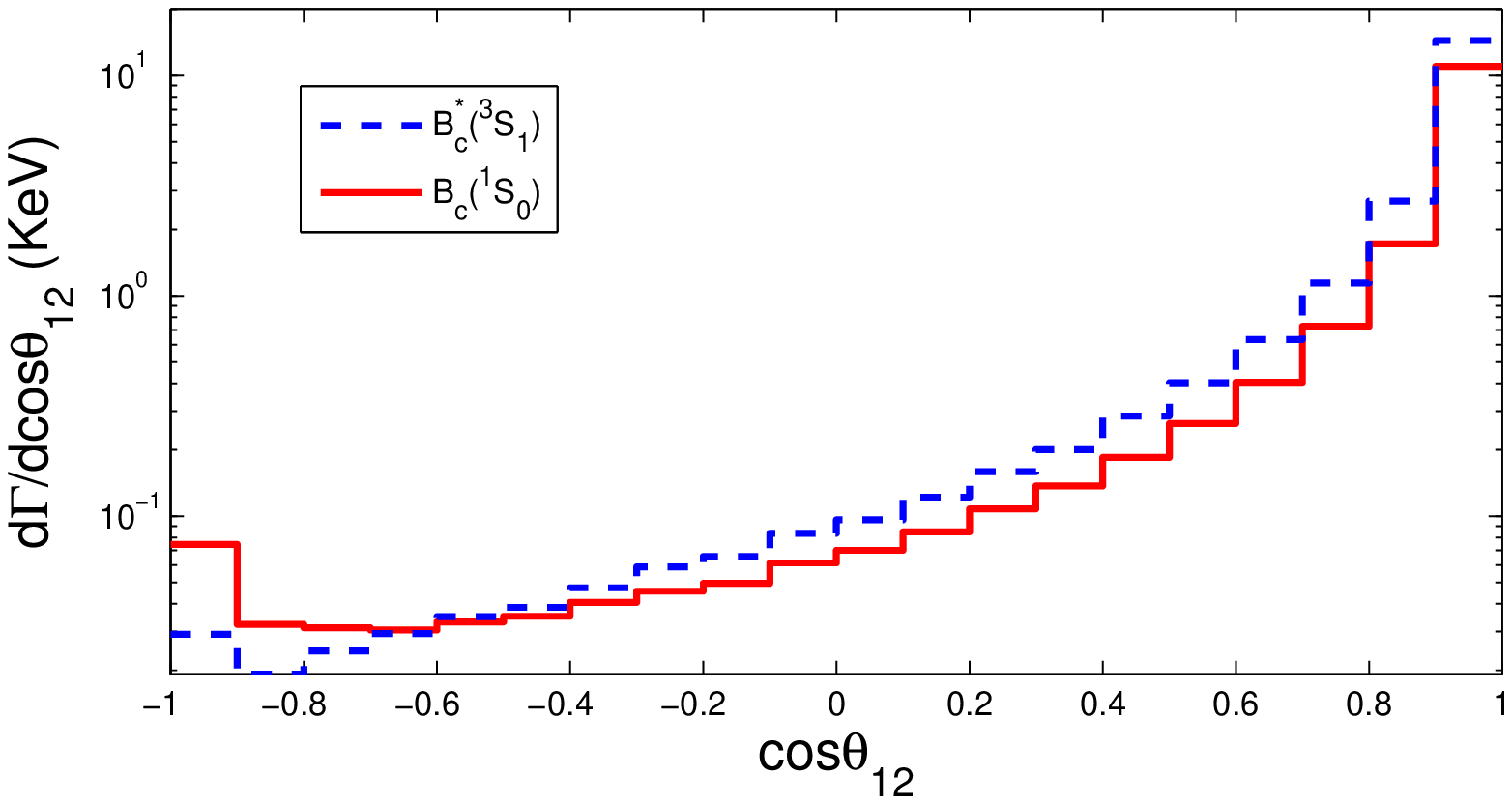}
\includegraphics[width=8cm]{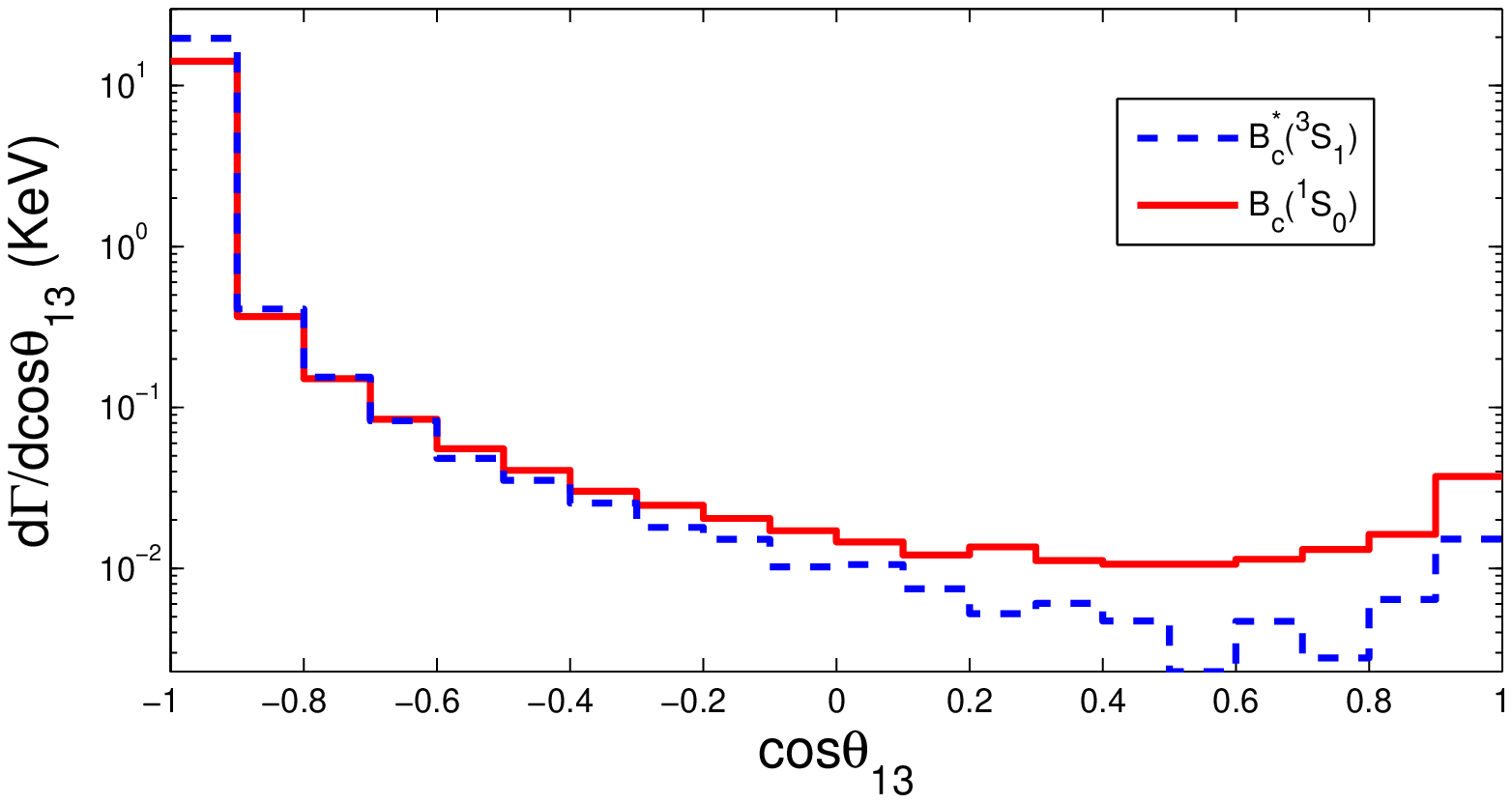}
\caption{\label{fig4} Differential decay widths $d\Gamma/dcos\theta_{12}$ (left) and $d\Gamma/dcos\theta_{13}$ (right)
for processes $H^0(k) \to B_c^{(*)}(p_0)+\bar{c}(p_5)+b(p_6)$, where $\theta_{12}$($\theta_{13}$) is the angle between $\protect\overrightarrow{p_0}$ and $\protect\overrightarrow{p_5}$($\protect\overrightarrow{p_6}$) in the Higgs boson rest frame.}
\end{figure}

To make our analysis more helpful to the experiments detection,
the differential distributions of invariant masses $s_1=(p_0+p_5)^2$ and $s_2=(p_5+p_6)^2$, i.e.
$d\Gamma/ds_1$ and $d\Gamma/ds_2$ are presented in Fig. \ref{fig3}.
And the differential distributions $d\Gamma/dcos\theta_{12}$ and $d\Gamma/dcos\theta_{13}$ are displayed
in Fig. \ref{fig4}, where $\theta_{12}$ is the angle between $\protect\overrightarrow{p_0}$ and $\protect\overrightarrow{p_5}$, and $\theta_{13}$ is that between $\protect\overrightarrow{p_0}$ and $\protect\overrightarrow{p_6}$ in the Higgs boson rest frame.
It can be found that the largest differential width is
achieved when $B^{(*)}_c$ mesons and $\bar{c}$ quark fly side by side ($\theta_{12}=0$), or $B^{(*)}_c$ mesons and $b$ quark fly back to back
($\theta_{13}=\pi$).
Here is the reason:
\begin{itemize}
  \item firstly, Feynman diagrams (2) and (4) in Fig. \ref{fig1} (i.e. the $Hb\bar{b}$ coupling one) dominate the processes;
  \item then, the gluon propagator $\frac{1}{(p_3+p_5)^2}$ in $M^{\prime}_{2,4}(n)$ reaches its peak value
when $B^{(*)}_c(p_0)$ mesons (in fact, the constitute quark $c(p_3)$) and $\bar{c}(p_5)$ quark go to the same direction.
\end{itemize}

Now, we will compare our results with the fragmentation ones.
To make it more clear, we divide our complete NRQCD leading-order results into three groups (only QCD contributions are included here):
contributions from $ Hb\bar{b} $ coupling (diagrams (2,4) in Fig. \ref{fig1});
contributions from $ Hc\bar{c} $ coupling (diagrams (1,3) in Fig. \ref{fig1});
the cross-terms of the previous two groups.
The corresponding branching fractions are displayed in Table \ref{tab5}, where the percentages in brackets are ratios of the contributions from the three groups relative to the total branching fractions.
Obviously, the contribution from $ Hb\bar{b} $ coupling dominates the processes, about two orders of magnitude bigger than the $ Hc\bar{c} $ one.
\begin{table}
\caption{\label{tab5} Branching fractions of $H^0(k) \to B_c^{(*)}(p_0)+\bar{c}(p_5)+b(p_6)$ through complete leading-order NRQCD calculation (QCD contributions only). The percentages in brackets are ratios of the contributions from the three groups relative to the total branching fractions.}
\begin{center}
  \begin{tabular}{c|c|c}
  \toprule
   & $B_c$ & $B^*_c$ \\
    \hline
    total branching fractions (QCD only) & $3.65 \times 10^{-4}$ & $4.96 \times 10^{-4}$ \\
    \hline
    $ Hb\bar{b} $ coupling & $3.61 \times 10^{-4}$(+98.9\%) & $4.97 \times 10^{-4}$(+100.2\%) \\
    $ Hc\bar{c} $ coupling & $8.82 \times 10^{-7}$(+0.2\%) & $7.32 \times 10^{-7}$(+0.2\%) \\
    cross-terms & $3.34 \times 10^{-6}$(+0.9\%) & $-1.85 \times 10^{-6}$(-0.4\%) \\
  \botrule
  \end{tabular}
\end{center}
\end{table}
\begin{table}
\caption{\label{tab6} Branching fractions of $H^0 \to B^{(*)}_c$ through $\bar{b}$ and $c$ quark fragmentation calculation.}
\begin{center}
  \begin{tabular}{c|c|c}
  \toprule
   & $B_c$ & $B^*_c$ \\
    \hline
    $H^0 \to b\bar{b} \to B^{(*)}_c$ & $3.10 \times 10^{-4}$ & $4.36 \times 10^{-4}$ \\
    $H^0 \to c\bar{c} \to B^{(*)}_c$ & $2.00 \times 10^{-7}$ & $1.73 \times 10^{-7}$ \\
  \botrule
  \end{tabular}
\end{center}
\end{table}

Under the fragmentation calculation, the branching fractions can be obtained by Higgs boson decays to $b\bar{b}$ or $c\bar{c}$,
following the $\bar b$ or $c$ quark fragments to $B^{(*)}_c$ mesons.
At $m_H=125.7GeV$, the decay rates of $H^0 \to b\bar{b}$ and $H^0 \to c\bar{c}$ are $56.6\%$ and $2.85\%$ respectively \cite{Heinemeyer:2013tqa}. As for the fragmentation probabilities of $\bar{b}/c \to B^{(*)}_c$,
by adopting the Eq.s (13-16) in Ref. \cite{Braaten:1993jn} but our input parameters,
we obtain the probabilities:
\begin{eqnarray}
P_{\bar{b} \to B_c}&=& 5.48\times10^{-4},\nonumber\\
P_{\bar{b} \to B^*_c}&=& 7.70\times10^{-4},\nonumber\\
P_{c \to B_c}&=& 7.02\times10^{-6},\nonumber\\
P_{c \to B^*_c}&=& 6.06\times10^{-6}. \label{11}
\end{eqnarray}
When multiplying these fragmentation probabilities by the decay rates of Higgs decays to $b\bar{b}/c\bar{c}$ couples,
the branching fractions of $H^0 \to B^{(*)}_c$ can be obtained directly,
which are presented in Table \ref{tab6}.
It is found that the fragmentation results consist with the complete leading-order calculation well\footnote{In fact, the results related to $ Hb\bar{b} $ coupling agree under the two framework, yet fragmentation results related to $ Hc\bar{c} $ coupling deviate by a factor of 4 from the complete leading-order calculation. We believe that the non-relativistic assumption ($ v \simeq 0 $) is not an apposite one when $c$ quark propagator is involved since $c$ quark is not heavy enough in the bound state. Although both framework have adopted the assumption, it might enlarge the differences of those two calculation methods when $c$ quark propagator emerges. We guess that the higher order correction in $v$ might reduce the discrepancy here. Luckily, the contribution containing the $ Hc\bar{c} $ coupling is small in comparison with the $ Hb\bar{b} $ coupling one here.}.

Finally, let's take a look at the branching fractions of Higgs boson decays to charmonium and bottomonium.
Within complete leading-order calculation, we obtain the branching fractions of $H^0 \to (\eta_c, J/\psi)+\bar{c}+c$ and
$H^0 \to (\eta_b, \Upsilon)+\bar{b}+b$ channels as follows (only QCD contributions are included):
\begin{eqnarray}
Br({H^0 \to \eta_c+\bar{c}+c})&=& 7.87\times10^{-5},\nonumber\\
Br({H^0 \to J/\psi+\bar{c}+c})&=& 7.59\times10^{-5},\nonumber\\
Br({H^0 \to \eta_b+\bar{b}+b})&=& 8.89\times10^{-5},\nonumber\\
Br({H^0 \to \Upsilon+\bar{b}+b})&=& 6.61\times10^{-5}.
\end{eqnarray}
Which are about a factor of 1/5 smaller than the $B_c$ one. Consistent results can also be obtained via the fragmentation calculation (fragmentation probabilities of $b/c$ quarks fragment into quarkonia can be found in Ref. \cite{Braaten:1993mp}).
It is worth noting that a factor 2 should be multiplied since both the quark and anti-quark
can fragment into the quarkonia. The $ Hb\bar{b} $ coupling, which appears in Higgs decays to $B_c^{(*)}$, is about
10 times larger than the $ Hc\bar{c} $ coupling, which
appears in the Higgs decays to charmonium. Hence, the Higgs branching
fraction to $B_c^{(*)}$ is larger than the Higgs branching fraction
to charmonium. The propagator $\frac{1}{(p_3+p_5)^2}$ in $M'_{2,4}(n)$
peaks at much larger values when the gluon fragments to $c\bar c$ than
when it fragments to $b\bar b$. This effect accounts for the
enhancement of the Higgs branching fraction to $B_c^{(*)}$ relative to
the Higgs branching fraction to bottomonium.

In contract to the exclusive radiative Higgs decay processes $H^0 \to (J/\psi, \Upsilon(nS)) + \gamma$, we find that the branching fractions of $H^0 \to B_c^{(*)}+\bar{c}+b$ are even larger. Latest results for $H^0 \to (J/\psi, \Upsilon(nS)) + \gamma$ can be found in Ref. \cite{Koenig:2015pha}, where direct amplitudes are evaluated at next-to-leading order in $ \alpha_s $ and loop-induced indirect amplitudes are also included. The branching fractions obtained there are $\sim 10^{-6}$ for $ J/\psi $ and $\sim 10^{-9}$ for $ \Upsilon(nS) $, which are two and five orders of magnitude smaller than the $ B_c^{(*)} $ one respectively.

\section{Summary and Conclusions}

In this work, we calculate the leading-order decay widths of processes $H^0 \to B^{(*)}_c+\bar{c}+b$ under NRQCD formulism,
which include both QCD and QED contributions. Corrections from triangle top quark loop and $HWW(HZZ)$ couplings are also discussed.
Uncertainties of the widths caused by the matrix elements $\langle {\cal O}[n]\rangle$, quark masses and running coupling constant $\alpha_s(\mu)$ are taken into consideration.
We also estimate the total branching fractions of $B^{(*)}_c$ mesons in Higgs boson decays, as well as the events at HL/HE-LHC.
Effects on the total branching fraction when the $Hb\bar{b}$ coupling deviates from the Standard Model is discussed.
To make the analysis more helpful, the differential distributions $d\Gamma/ds_{1,2}$ and $d\Gamma/dcos\theta_{12,13}$ are presented.
Moreover, a comparison between our results and the ones calculated under fragmentation formulism is displayed in detail.
And the comparison between the branching fractions of Higss boson decays to $B^{(*)}_c$ and
those of Higgs decays to charmonium ($\eta_c, J/\psi$) or bottomonium ($\eta_b, \Upsilon$) is also presented.

We find that QCD contribution dominates the processes as is expected, and the varying $c$ quark mass has a strong influence on the total branching fraction (the correction to the total branching rates can reach about $25\%$). We also find that Feynman diagrams with the coupling of $Hb\bar{b}$ (i.e., Fig.\ref{fig1} (2,4)) dominate, contributions from the triangle top quark loop and other couplings ($Hc\bar{c}$, $HWW$ and $HZZ$) are small comparatively. Differential distributions $d\Gamma/dcos\theta_{12,13}$ show that the largest differential width
emerges when $B^{(*)}_c$ mesons and $\bar{c}$ jet fly side by side ,
or $B^{(*)}_c$ mesons and $b$ jet fly back to back at the Higgs boson rest frame.
Moreover, the calculation results inform that
the branching fractions of Higgs boson decays to $B^{(*)}_c$ are bigger than those of Higgs decays to quarkonia.

According to our study, when $14~TeV$ LHC delivering the integrated luminosity of $3~ab^{-1}$ (i.e. HL-LHC),
about $1.4\times10^5$ $B_c$ events can be produced through Higgs boson decays. Considering that the detection efficiency of fully constructed channel $B_c \to J/\psi(\to l^+l^-)+\pi^+$ is around six in ten thousands, about 84 $B_c$ candidates can be observed at HL-LHC, and 3.5 times bigger when the center-of-mass energy is upgraded to $33~TeV$ (i.e. HE-LHC).

As a final remark, the study of $B_c$ production in Higgs boson decays could be considered as the choice for the measurement of $ Hb\bar{b} $ coupling  at future HL/HE-LHC. And it is also a place searching for signals which deviate from the Standard Model.

\vspace{.7cm} {\bf Acknowledgments}

This work was supported in part by Ministry of Science and
Technology of the People's Republic of China (2015CB856703), and by the
National Natural Science Foundation of China (NSFC) under the grant 11375200.


\clearpage


\begin{thebibliography}{99}


\bibitem{Aad:2012tfa}
  G.~Aad {\it et al.} [ATLAS Collaboration],
  ``Observation of a new particle in the search for the Standard Model Higgs boson with the ATLAS detector at the LHC,''
  Phys.\ Lett.\ B {\bf 716}, 1 (2012)
  [arXiv:1207.7214 [hep-ex]].

\bibitem{Chatrchyan:2012xdj}
  S.~Chatrchyan {\it et al.} [CMS Collaboration],
  ``Observation of a new boson at a mass of 125 GeV with the CMS experiment at the LHC,''
  Phys.\ Lett.\ B {\bf 716}, 30 (2012)
  [arXiv:1207.7235 [hep-ex]].

\bibitem{Orestano:2015ljx}
  D.~Orestano [ATLAS Collaboration],
  ``Latest Higgs physics results with the ATLAS detector,''
  Int.\ J.\ Mod.\ Phys.\ Conf.\ Ser.\  {\bf 39}, 1560091 (2015),
  doi:10.1142/S2010194515600915.

\bibitem{Mochizuki:2015wua}
  K.~Mochizuki [ATLAS Collaboration],
  ``Search for the Higgs boson in fermionic channels using ATLAS detector,''
  J.\ Phys.\ Conf.\ Ser.\  {\bf 623}, no. 1, 012020 (2015),
  doi:10.1088/1742-6596/623/1/012020.

\bibitem{Flechl:2015xxa}
  M.~Flechl [CMS Collaboration],
  ``Higgs boson discovery and recent results,''
  arXiv:1510.01924 [hep-ex].

\bibitem{ac2015}
  The ATLAS and CMS Collaborations,
  ``Measurements of the Higgs boson production and decay rates and constraints on its couplings from a combined ATLAS and CMS analysis of the LHC pp collision data at $\sqrt{s}$ = 7 and 8 TeV,''
  ATLAS-CONF-2015-044, CMS-PAS-HIG-15-002.

\bibitem{Kourkoumelis:2015vza}
  C.~Kourkoumelis [ATLAS and CMS Collaborations],
  ``ATLAS and CMS review of SM Higgs boson searches: bosonic and fermionic decays,''
  PoS Charged {\bf 2014}, 001 (2015);\\
  P.~Je$\check{z}$ [ATLAS and CMS Collaborations],
  ``Review of SM Higgs properties measured by ATLAS and CMS: couplings, spin, mass,''
  PoS Charged {\bf 2014}, 002 (2015).

\bibitem{Todesco:2013cca}
  E.~Todesco, M.~Lamont and L.~Rossi,
  ``High luminosity LHC and high energy LHC,''
  Conference: C12-05-28.4.

\bibitem{higgscs}
  LHC Higgs Cross Section Working Group,
  https://twiki.cern.ch/twiki/bin/view/LHCPhysics\\/HiggsEuropeanStrategy\#SM\_Higgs\_production\_cross\_se\_AN2.

\bibitem{CEPC-SPPCStudyGroup:2015csa}
  [CEPC-SPPC Study Group Collaboration],
  ``CEPC-SPPC Preliminary Conceptual Design Report. 1. Physics and Detector,''
  IHEP-CEPC-DR-2015-01, IHEP-TH-2015-01, HEP-EP-2015-01;
  ``CEPC-SPPC Preliminary Conceptual Design Report. 2.  Accelerator,''
  IHEP-CEPC-DR-2015-01, IHEP-AC-2015-01.

\bibitem{Aad:2015sda}
  G.~Aad {\it et al.} [ATLAS Collaboration],
  ``Search for Higgs and Z Boson Decays to $J/\psi \gamma$ and $\Upsilon(nS)\gamma$ with the ATLAS Detector,''
  Phys.\ Rev.\ Lett.\  {\bf 114}, no. 12, 121801 (2015)
  [arXiv:1501.03276 [hep-ex]].

\bibitem{Achasov:1991ms}
  N.~N.~Achasov and V.~K.~Besprozvannykh,
  ``Decays $\psi, \Upsilon \to H(a) \gamma$ and $H \to \psi \gamma, \Upsilon \gamma$,''
  Sov.\ J.\ Nucl.\ Phys.\  {\bf 55}, 1072 (1992)
  [Yad.\ Fiz.\  {\bf 55}, 1934 (1992)].

\bibitem{Bodwin:2013gca}
  G.~T.~Bodwin, F.~Petriello, S.~Stoynev and M.~Velasco,
  ``Higgs boson decays to quarkonia and the $H\bar{c}c$  coupling,''
  Phys.\ Rev.\ D {\bf 88}, no. 5, 053003 (2013)
  doi:10.1103/PhysRevD.88.053003
  [arXiv:1306.5770 [hep-ph]];\\
  G.~T.~Bodwin, H.~S.~Chung, J.~H.~Ee, J.~Lee and F.~Petriello,
  ``Relativistic corrections to Higgs boson decays to quarkonia,''
  Phys.\ Rev.\ D {\bf 90}, no. 11, 113010 (2014)
  [arXiv:1407.6695 [hep-ph]].

\bibitem{Koenig:2015pha}
  M.~König and M.~Neubert,
  ``Exclusive Radiative Higgs Decays as Probes of Light-Quark Yukawa Couplings,''
  JHEP {\bf 1508}, 012 (2015),
  [arXiv:1505.03870 [hep-ph]].

\bibitem{Qiao:1998kv}
  C.~F.~Qiao, F.~Yuan and K.~T.~Chao,
  ``Quarkonium production in SM Higgs decays,''
  J.\ Phys.\ G {\bf 24}, 1219 (1998)
  [hep-ph/9805431].

\bibitem{Bodwin:1994jh}
  G.~T.~Bodwin, E.~Braaten and G.~P.~Lepage,
  ``Rigorous QCD analysis of inclusive annihilation and production of heavy quarkonium,''
  Phys.\ Rev.\ D {\bf 51}, 1125 (1995)
  [Phys.\ Rev.\ D {\bf 55}, 5853 (1997)]
  [hep-ph/9407339].

\bibitem{Petrelli:1997ge}
  A.~Petrelli, M.~Cacciari, M.~Greco, F.~Maltoni and M.~L.~Mangano,
 ``NLO production and decay of quarkonium,''
  Nucl.\ Phys.\ B {\bf 514}, 245 (1998)
  [hep-ph/9707223].

\bibitem{Lepage:1980fj}
  G.~P.~Lepage and S.~J.~Brodsky,
  ``Exclusive Processes in Perturbative Quantum Chromodynamics,''
  Phys.\ Rev.\ D {\bf 22}, 2157 (1980).

\bibitem{Chernyak:1983ej}
  V.~L.~Chernyak and A.~R.~Zhitnitsky,
  ``Asymptotic Behavior of Exclusive Processes in QCD,''
  Phys.\ Rept.\  {\bf 112}, 173 (1984).

\bibitem{Gao:2014xlv}
  D.~N.~Gao,
  ``A note on Higgs decays into Z boson and $J/\psi(\Upsilon)$,''
  Phys.\ Lett.\ B {\bf 737}, 366 (2014)
  [arXiv:1406.7102 [hep-ph]].

\bibitem{Kartvelishvili:2008tz}
  V.~Kartvelishvili, A.~V.~Luchinsky and A.~A.~Novoselov,
  ``Double vector quarkonia production in exclusive Higgs boson decays,''
  Phys.\ Rev.\ D {\bf 79}, 114015 (2009)
  [arXiv:0810.0953 [hep-ph]];\\
  V.~G.~Kartvelishvili, A.~V.~Luchinsky and A.~A.~Novoselov,
  ``Double production of vector quarkonia in exclusive Higgs boson decays,''
  Phys.\ Atom.\ Nucl.\  {\bf 73}, 949 (2010)
  [Yad.\ Fiz.\  {\bf 73}, 983 (2010)].

\bibitem{Qiao:2011yk}
  C.~F.~Qiao, L.~P.~Sun, D.~S.~Yang and R.~L.~Zhu,
  ``W Boson Inclusive Decays to Quarkonium at the LHC,''
  Eur.\ Phys.\ J.\ C {\bf 71}, 1766 (2011)
  [arXiv:1103.1106 [hep-ph]].

\bibitem{Liao:2011kd}
  Q.~L.~Liao, X.~G.~Wu, J.~Jiang, Z.~Yang and Z.~Y.~Fang,
  ``Heavy Quarkonium Production at LHC through $W$ Boson Decays,''
  Phys.\ Rev.\ D {\bf 85}, 014032 (2012)
  [arXiv:1111.4609 [hep-ph]];\\
  Q.~L.~Liao, X.~G.~Wu, J.~Jiang, Z.~Yang, Z.~Y.~Fang and J.~W.~Zhang,
  ``Excited Heavy Quarkonium Production at the LHC through $W$-Boson Decays,''
  Phys.\ Rev.\ D {\bf 86}, 014031 (2012)
  [arXiv:1204.2594 [hep-ph]].

\bibitem{Qiao:1996rd}
  C.~F.~Qiao, C.~S.~Li and K.~T.~Chao,
  ``Top quark decays into heavy quark mesons,''
  Phys.\ Rev.\ D {\bf 54}, 5606 (1996)
  [hep-ph/9603275].

\bibitem{Chang:2007si}
  C.~H.~Chang, J.~X.~Wang and X.~G.~Wu,
  ``Production of $B_c$ or $\bar{B}_c$ meson and its excited states via $\bar{t}$ quark or $t$ quark decays,''
  Phys.\ Rev.\ D {\bf 77}, 014022 (2008)
  [arXiv:0711.1898 [hep-ph]].

\bibitem{Sun:2010rw}
  P.~Sun, L.~P.~Sun and C.~F.~Qiao,
  ``The Next-to-Leading Order Corrections to Top Quark Decays to Heavy Quarkonia,''
  Phys.\ Rev.\ D {\bf 81}, 114035 (2010)
  [arXiv:1003.5360 [hep-ph]].

\bibitem{Chang:1992bb}
  C.~H.~Chang and Y.~Q.~Chen,
  ``The Production of $B_c$ or $\bar{B}_c$ meson associated with two heavy quark jets in Z0 boson decay,''
  Phys.\ Rev.\ D {\bf 46}, 3845 (1992)
  [Phys.\ Rev.\ D {\bf 50}, 6013 (1994)].

\bibitem{Deng:2010aq}
  L.~C.~Deng, X.~G.~Wu, Z.~Yang, Z.~Y.~Fang and Q.~L.~Liao,
  ``$Z_0$ Boson Decays to $B^{(*)}_c$ Meson and Its Uncertainties,''
  Eur.\ Phys.\ J.\ C {\bf 70}, 113 (2010)
  [arXiv:1009.1453 [hep-ph]];\\
  Z.~Yang, X.~G.~Wu, L.~C.~Deng, J.~W.~Zhang and G.~Chen,
  ``Production of the $P$-Wave Excited $B_c$-States through the $Z^0$ Boson Decays,''
  Eur.\ Phys.\ J.\ C {\bf 71}, 1563 (2011)
  [arXiv:1011.5961 [hep-ph]].

\bibitem{Qiao:2011zc}
  C.~F.~Qiao, L.~P.~Sun and R.~L.~Zhu,
  ``The NLO QCD Corrections to $B_c$ Meson Production in $Z^0$ Decays,''
  JHEP {\bf 1108}, 131 (2011)
  [arXiv:1104.5587 [hep-ph]].

\bibitem{Jiang:2015jma}
  J.~Jiang, L.~B.~Chen and C.~F.~Qiao,
  ``QCD NLO corrections to inclusive $B_c^*$ production in $Z^0$ decays,''
  Phys.\ Rev.\ D {\bf 91}, no. 3, 034033 (2015)
  [arXiv:1501.00338 [hep-ph]].



\bibitem{Eichten:1995ch}
  E.~J.~Eichten and C.~Quigg,
  ``Quarkonium wave functions at the origin,''
  Phys.\ Rev.\ D {\bf 52}, 1726 (1995)
  [hep-ph/9503356].

\bibitem{Hahn:2000kx}
  T.~Hahn,
  ``Generating Feynman diagrams and amplitudes with FeynArts 3,''
  Comput.\ Phys.\ Commun.\  {\bf 140}, 418 (2001)
  [hep-ph/0012260].

\bibitem{Bodwin:2002hg}
  G.~T.~Bodwin and A.~Petrelli,
  ``Order-$v^4$ corrections to $S$-wave quarkonium decay,''
  Phys.\ Rev.\ D {\bf 66}, 094011 (2002)
  [Phys.\ Rev.\ D {\bf 87}, no. 3, 039902 (2013)]
  [arXiv:1301.1079 [hep-ph]].

\bibitem{Mertig:1990an}
  R.~Mertig, M.~Bohm and A.~Denner,
  ``FEYN CALC: Computer algebraic calculation of Feynman amplitudes,''
  Comput.\ Phys.\ Commun.\  {\bf 64}, 345 (1991).

\bibitem{Hahn:1998yk}
  T.~Hahn and M.~Perez-Victoria,
  ``Automatized one loop calculations in four-dimensions and D-dimensions,''
  Comput.\ Phys.\ Commun.\  {\bf 118}, 153 (1999)
  [hep-ph/9807565].



\bibitem{Agashe:2014kda}
  K.~A.~Olive {\it et al.} [Particle Data Group Collaboration],
  ``Review of Particle Physics,''
  Chin.\ Phys.\ C {\bf 38}, 090001 (2014).

\bibitem{Heinemeyer:2013tqa}
  S.~Heinemeyer {\it et al.} [LHC Higgs Cross Section Working Group Collaboration],
  ``Handbook of LHC Higgs Cross Sections: 3. Higgs Properties,''
  arXiv:1307.1347 [hep-ph].

\bibitem{Chang:1992pt}
  C.~H.~Chang and Y.~Q.~Chen,
  ``The Decays of $B_c$ meson,''
  Phys.\ Rev.\ D {\bf 49}, 3399 (1994).

\bibitem{Braaten:1993jn}
  E.~Braaten, K.~m.~Cheung and T.~C.~Yuan,
  ``Perturbative QCD fragmentation functions for $B_c$ and $B_{c}$ * production,''
  Phys.\ Rev.\ D {\bf 48}, 5049 (1993)
  [hep-ph/9305206].

\bibitem{Braaten:1993mp}
  E.~Braaten, K.~m.~Cheung and T.~C.~Yuan,
  ``$Z^0$ decay into charmonium via charm quark fragmentation,''
  Phys.\ Rev.\ D {\bf 48}, 4230 (1993)
  [hep-ph/9302307].


\end{thebibliography}
\end{document}